\documentclass[aps,prl,twocolumn,showpacs,groupedaddress,floatfix,superscriptaddress]{revtex4}
\usepackage{graphicx,graphics}

\usepackage{color}
\usepackage{bm}
\begin{document}

\title{Direct measurements of hydrophobic slippage using double-focus fluorescence cross-correlation. }

\author{Olga~I.~Vinogradova}
\email[Corresponding author: ] {oivinograd@yahoo.com}

\affiliation{A.N.~Frumkin Institute of Physical Chemistry and
Electrochemistry, Russian Academy of Sciences, 31 Leninsky
Prospect, 119991 Moscow, Russia} \affiliation{CNRS UMR 7636,
ESPCI, 10 rue Vauquelin, 75231 Paris Cedex 05, France}
\affiliation{Institut f\"ur Technische Chemie und Makromolekulare
Chemie,  RWTH Aachen, Pauwelsstr. 8, 52056 Aachen, Germany}

\author{Kaloian Koynov}
\affiliation{Max Planck Institute for Polymer Research, Ackermannweg
10, 55128 Mainz, Germany}

\author{Andreas Best}
\affiliation{Max Planck Institute for Polymer Research, Ackermannweg
10, 55128 Mainz, Germany}

\author{Fran\c{c}ois Feuillebois}
\affiliation{CNRS UMR  7636, ESPCI, 10 rue Vauquelin, 75231 Paris
Cedex 05, France}

\date{\today}

\begin{abstract}
We report results of direct measurements of velocity profiles in a
microchannel with hydrophobic and hydrophilic walls, using a new
high precision method of double-focus spacial fluorescence
cross-correlation under a confocal microscope. In the vicinity of
both walls the measured velocity profiles do not turn to zero by
giving a plateau of constant velocity. This apparent slip is
proven to be due to a Taylor dispersion, an augmented by shear
diffusion of nanotracers in the direction of flow. Comparing the
velocity profiles near the hydrophobic and hydrophilic walls for
various conditions shows that there is a true slip length due to
hydrophobicity. This length, of the order of several tens of
nanometers, is independent on electrolyte concentration and shear
rate.

\end{abstract}


\maketitle


For more than hundred years scientists and engineers have assumed
and successfully applied no-slip boundary conditions to model
experiments in fluid mechanics~\cite{lamb.h:1932}. However, it has
been recently well recognized, that the success of this famous
no-slip postulate reflected mostly a macroscopic character and
insensitivity of old experiments. Reducing the size of
investigated systems to micro- and, especially, nanodimension led
to a very definite conclusion that the no-slip condition does not
always apply~\cite{vinogradova:99}. It is now clear that many
systems should allow for an amount of slippage, described in terms
of a slip length: $v_s = b \partial_z v,$ where $v_s$ is the slip
(tangential) velocity at the wall and the axis $z$ is normal to
the surface. What, however, remains a matter of active debates is
the amplitude of slip, and its variation with interfacial
properties and parameters of the flow.

From the theoretical point of view the situation is reasonably
clear. Slippage should not appear on a hydrophilic surface, except
as at very high shear rate~\cite{thompson.pa:1997}. A slip length
of the order of hundred nanometers or smaller is, however,
expected for a hydrophobic
surface~\cite{vinogradova:95,barrat:99,andrienko.d:2003}. On the
experimental side, no consensus is achieved so far. While some
experimental data are consistent with the theoretical expectations
both for
hydrophilic~\cite{vinogradova:03,charlaix.e:2005,joly.l:2006,vinogradova.oi:2006,honig.cdf:2007}
and hydrophobic
surfaces~\cite{vinogradova:03,charlaix.e:2005,joly.l:2006,huang.p:2006},
some other reports completely escape from this picture with both
quantitative (slippage over hydrophilic surface, shear rate
dependent slippage, rate threshold for slip, etc) and quantitative
(slip length of several $\mu$ms) discrepancies (for a recent
review see~\cite{lauga.e:2005}). Clearly, in order to rationalize
the experimental situation, new data are necessary. These data
should preferably be obtained with a new experimental technique.

Basically, two types of experimental methods were used to study
boundary conditions. High-speed force measurements performed with
the surface forces apparatus (SFA)~\cite{horn:00,charlaix.e:2005}
or atomic force microscope (AFM)~\cite{vinogradova:03} allows to
deduce a drag force, with the subsequent comparison with a theory
of a film drainage~\cite{vinogradova:95}. This approach, being
extremely accurate at the nanoscale, does not provide
visualization of the flow profile, so that this type of
measurements should be identified as indirect. Direct approaches
to flow profiling, or a velocimetry, take advantage of various
optics to monitor tracer
particles~\cite{pit:2000,tretheway.dc:2002,josef.p:2005}. Their
accuracy is normally much lower than that of force methods due to
relatively low optical resolution, system noise due to
polydispersity of tracers, and difficulties in decoupling of
directed flow from diffusion. As a consequence, it is normally
expected that a slippage of the order of a few tens nanometers
cannot be detected by a velocimetry technique.

In this Letter we report direct high-precision measurements at the
nanoscale performed with a new optical technique. As an
alternative to the existing FTIR~\cite{pit:2000},
$\mu$-PIV~\cite{tretheway.dc:2002,josef.p:2005}, and
TIRV~\cite{huang.p:2006} methods we here use a recently suggested
technique, based on a double-focus spatial fluorescence
cross-correlation (DF FCS)~\cite{lumma.d:2003}. Our method allows
one to use much smaller tracers, an order of magnitude higher
shear rates, and to get orders of magnitude better statistics as
compared with the state of the art. These give us at least an
order of magnitude improvement in accuracy compared with other
velocimetry approaches. Results obtained for various experimental
conditions allow us to deduce the true hydrophobic slip length,
which is proven to be of the order of several tens of nanometers,
and is independent on electrolyte concentration and shear rate.

\begin{figure}
\includegraphics[width=5cm,clip]{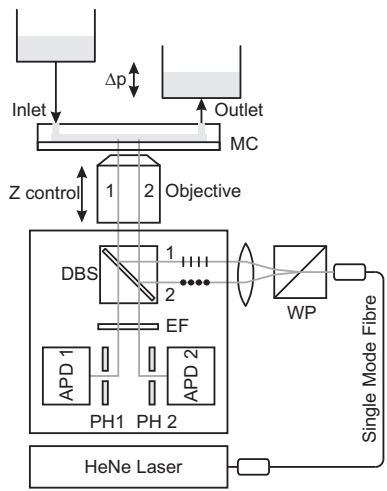}
\caption{Schematic of the experimental setup. Abbreviations are: MC
- Microchannel, WP - Wollaston Prism,  DBS - Dichroic Beamsplitter,
EF - Emission Filter, APD - Avalanche Photodiode, and PH - Pinhole.}
\label{fig1}
\end{figure}

Our microchanel  was formed by a three-layer sandwich
construction. The lowest layer was a standard microscope cover
slide made of borosilicate glass with a thickness of 170 $\mu$m, a
root-mean-square roughness of the range 1-2 nm. The water
advancing contact angle on this slide was measured to be below
$5^{\circ}$. The channel itself was created by cutting out a hole
in an adhesive polymer film (Tessa, Germany) with a thickness of
around 100 $\mu$m, that forms the smallest dimension of the
channel, directed along the $z$ axis. The channel extension along
the $y$ axis (along the wall and perpendicular to the flow
direction) is about 1.5 mm. Finally, the top layer was formed by a
1-mm-thick cover glass. Its surface was made hydrophobic by
silanization and the water contact angle was measured to be
$85-90^{\circ}$. An optically transparent polycarbonate block
served as a support and for connection of the chamber to the
external flow system. A hydrostatic pressure gradient was created
by a system of two beakers at different heights, which allowed us
to vary a shear rate near the wall in the range $\lambda=800-3000$
s$^{-1}$. As a tracers we used fluorescently labeled latex
spheres, carboxylate-modified FluoSpheres 580/605 (Molecular
Probes, Eugene, Oregon). The particles had a radius of $R\sim20$
nm and a polydispersity of about 20\%. Experiments were carried
out in water and NaCl aqueous solutions with concentrations in the
range between $10^{-5}$ mol/L and $10^{-2}$ mol/L.

\begin{figure}
\includegraphics[width=4cm,clip]{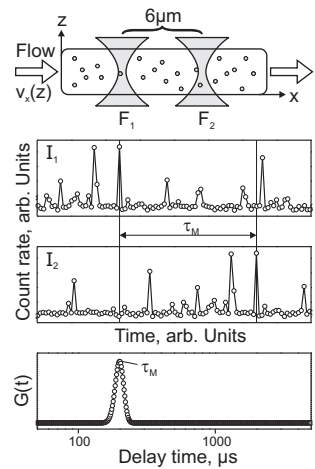}
\caption{Schematics of the basic idea of the double-focus spatial
fluorescence cross-correlation method. Two laser foci are placed
along the $x$ axis separated by a distance of 6 $\mu$m. They
independently record the time-resolved fluorescence intensities
$I_1 (t)$ and $I_2 (t)$. The forward cross-correlation of these
two signals yelds $G(t)$. Two foci are scanned simultaneously
along the $z$ axis to probe the velocity profile $v(z)$.}
\label{fig2}
\end{figure}

The scheme of the DF FSC method is shown on Fig~\ref{fig1} and is
described in details before~\cite{lumma.d:2003}. Briefly, we used
a commercial FCS setup (Carl Zeiss Jena, Germany) consisting of
the module ConfoCor 2 and the inverted microscope model Axiovert
200. For the present experiments, we employed a water immersion
objective (Zeiss, C-Apochromat $40\times$, NA 1.2). The optical
system was modified so that an external laser beam could be
coupled into the confocal optics. For fluorescence excitation, the
543-nm line of a 1-mW helium-neon laser was used. The laser beam
was split by means of a Wollaston prism. Behind the prism, the two
beams are polarized perpendicularly to each other and exhibit an
angular separation of $0.5^{\circ}$. After passing through two
additional lenses these beams are fed into the confocal
microscope. Our alignments result in two optically equivalent,
almost diffraction-limited laser foci (diameter $\sim 400$ nm,
height $\sim 3$ $\mu$m) separated by a distance of $6.0\pm0.1$
$\mu$m in object space as is schematically shown in
Fig.~\ref{fig2}. As the fluorescence tracers are flowing along the
channel they are crossing consecutively the two foci, producing
two time-resolved fluorescence intensities $I_1(t)$ and $I_2(t)$
recorded independently from the avalanche photo-diodes APD~1 and
APD~2. The time cross-correlation function of $I_1(t)$ and
$I_2(t)$ can be calculated as $G(\tau)=\langle I_1(t)I_2(t+\tau)
\rangle_t / \langle I_1(t) \rangle_t \langle I_2(t) \rangle_t$ and
typically exhibit a local maximum. The position of this maximum
$\tau_{\rm M}$ is characteristic of the local velocity of the
tracers.

To determine the velocity profile we have scanned the foci
position across the channel. At each $z$ position, a series of 10
independent data acquisitions was carried out. The acquisition
time was either 30 s or 60 s, necessitating longer measurements
close to the channel walls, where small flow velocities are found.
Indeed, consider the worst case when the foci are centered on the
wall. Since the concentration in particles is about 1 per
femtoliter, the number of particles carried by the shear flow
which enter the focus half-elliptical area during 60 s is about
$N=6\times 10^4$. This gives a satisfactory signal to noise ratio
$\sqrt{N}$ larger than $10^2$.  The independent cross-correlation
functions acquired at position $z$ were fitted with a Gaussian
function for more precise determination of $\tau_{\rm M}(z)$,
yielding the particle velocity $v(z)=\Delta s/ \tau_{\rm M}(z)$,
where $\Delta s$ is the distance between foci.  For every salt
concentration we have repeated experiments several times with
freshly prepared channels and varied the pressure gradient.

\begin{figure}
\includegraphics[width=6.5cm,clip]{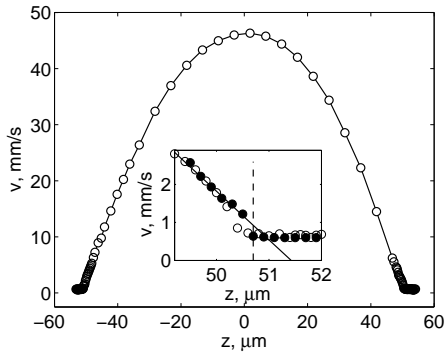}
\caption{Typical velocity profile $v(z)$ measured in a $\sim 100$
$\mu$m channel with $10^{-4}$ mol/L NaCl solution. Inset: The
observed velocity profile in the vicinity of the wall and
schematics of the procedure for a determination of the apparent
slip length ($b_{\rm app} \sim 445$ nm) and the wall location
($z=50.7 \mu$m). } \label{fig3}
\end{figure}

\begin{figure}
\includegraphics[width=6.5cm,clip]{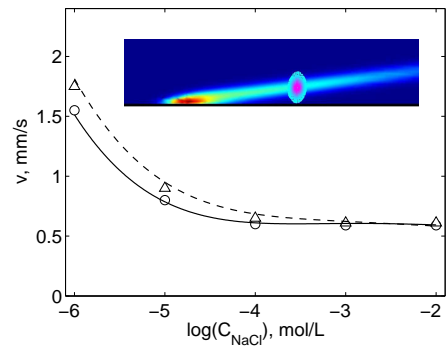}
\caption{The apparent velocity at the plateau region, $v_{\rm
app}$, at the hydrophobic wall (triangles) and hydrophilic wall
(circles) as a function of concentration of NaCl. Concentration
$10^{-6}$ mol/L corresponds to a case of pure water. Dashed and
solid curves show the values predicted for $b=100$ nm and $b=0$,
correspondingly ($\lambda = 1750$ s$^{-1}$).
 Inset
shows typical calculated isoconcentration lines
($\overline{c}(x,z)=$Ct) distorted by dispersion together with
isolines of light intensity profile at the downstream focus
($\overline{i_2}(x,z)=$Ct). } \label{plateau}
\end{figure}

A typical measured velocity profile is shown in Fig.~\ref{fig3}.
As expected, the central region, where the velocity of the tracer
particles reflect that a liquid~\cite{lumma.d:2003}, the profile
exhibits the parabolic shape predicted by the classical theory.
However, when foci presumably enter the wall, this parabola does
not turn to zero by giving a plateau of non-zero constant
velocity. This is observed for both hydrophobic and hydrophilic
surfaces. The apparent velocity at the plateau region is always
higher for a hydrophobic surface and decreases with added salt and
as shown in Fig.~\ref{plateau}.

\begin{figure}
\includegraphics*[width=5.5cm,clip]{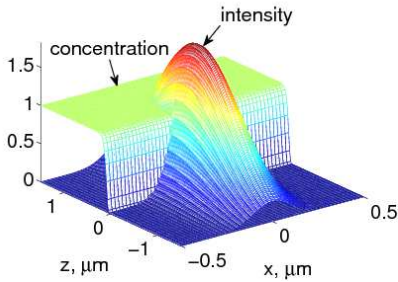}
\caption{Concentration at upstream focus from balance of Brownian
motion and electrostatic repulsive forces for the typical case of
$10^{-4}$ mol/L NaCl, and intensity at the upstream focus when
centered at 100 nm from the wall. The concentration of illuminated
particles is the product of these quantities.}
\label{fig5}
\end{figure}

The apparent velocity at the wall is too large to reflect the true
liquid slippage over it. Earlier estimates~\cite{lumma.d:2003}
suggested that in the vicinity of the wall the tracers are
submitted to a Taylor dispersion, e.g. their diffusion is
augmented by shear, enhancing a migration speed in the direction
of flow. Now we model this effect precisely. Like
in~\cite{lumma.d:2003} we assume an ergodic system and interpret
the time cross-correlation $G(\tau)$ as:
\begin{equation}
G(\tau) = \frac{ \int \!\! \int i_1(\bm{r}) \, i_2(\bm{r}\,') \,
\Phi(\bm{r},\bm{r}\,',\tau) \, \mbox{d}^3\bm{r} \,
\mbox{d}^3\bm{r}\,'} {\overline{C}^2 \,  \int \!\! \int
i_1(\bm{r}) \, i_2(\bm{r}\,') \, \mbox{d}^3\bm{r} \,
\mbox{d}^3\bm{r}\,'} \label{eq:G_from_phi}
\end{equation}
where $\bm{r}=(x,y,z)$, $\bm{r}\,'=(x',y',z')$ and the average
concentration of labelled particles is denoted by $\overline{C}$.
The real-space detection efficiencies $i_1(\bm{r}\,)$ and
$i_2(\bm{r}\,)$ for focus 1 and 2 were given as ellipsoidal
Gaussian functions. The function $\Phi$ is given as the solution
of the advection-diffusion equation from a point source in a flow
field with uniform velocity. In the vicinity of the wall, the
particles are repelled by an electrostatic force, $F$, so that
they do not fill up completely the ellipsoidal Gaussian lightened
region. Keeping the same notation for the light intensity, we thus
replace $i_1(\bm{r}\,)$ by $i_1(\bm{r}\,)c_e(\bm{r}\,)$, where
$c_e(\bm{r}\,)$ is the particle equilibrium concentration profile
at the upstream focus, to be detailed below. We also take into
account the velocity gradient in the advection flows, thereby
introducing the mechanism for Taylor dispersion. The
advection-diffusion equation to be solved is:
\begin{equation}
\partial c_\tau +(\lambda z+b)\,\partial c_x + \partial (w c)_z =
D \nabla^2 c \label{advection-diffusion}
\end{equation}
where $w=F/(6\pi R \mu)$ is the migration velocity of a particle
along $z$ in a fluid of viscosity $\mu$ and $D$ is the Einstein
diffusion coefficient. Here we neglect the hydrodynamic
interactions between the particles and wall. Since the distance
from the wall $h \gg R$, the energy of electrostatic interaction
of a particle with the wall is $U=q \phi_1 \exp (-\kappa h),$ i.e.
the particle is considered as a point charge $q$ [Correspondingly,
$F=-dU/dh$]. Here $\phi_1$ is the surface potential of the wall at
the given concentration of electrolyte with an inverse Debye
length $\kappa$.  The charge is given by $q =4 \pi R^2 q_s = 4 \pi
R \epsilon_0 \epsilon \phi_2 (\kappa R +
    1),$ where $q_s$ is the surface charge density, and $\phi_2$ is the
surface potential of the particle. Then, according to the
Boltzmann law, the equilibrium concentration of particles in the
vicinity of the wall is $c_e(\bm{r}\,)= c_0 \exp (-U/k_B T)=c_0
\exp (-A \exp (-\kappa h))$, where $A=4 \pi R  \phi_1 \phi_2
\epsilon_0 \epsilon (1 + \kappa R) / (k_B T)$. For $\phi_1$ we
used data of~\cite{horn.rg:1989,ducker.wa:1992}, data for $\phi_2$
were smaller according to electrokinetic measurements. With these
parameters the values of $A$ were of the order of 25. Instead of
solving for an initial point source, multiplying by
$i_1(\bm{r}\,)c_1(\bm{r}\,)$ and calculating the integral on
$\bm{r}$ like in the numerator of Eq.\ref{eq:G_from_phi}, we may
from linearity of Eq.\ref{advection-diffusion} solve for the
initial cloud of illuminated particles with concentration
$i_1(\bm{r}\,)c_1(\bm{r}\,)$. The result $c(\bm{r},\tau)$ is then
multiplied by $i_2(\bm{r}\,')$ and integrated. This integral goes
through a maximum at some time $\tau_M$, which is interpreted as
the transit time between foci, like in Fig.~\ref{fig2}. Typical
distributions of the concentration $c_e$ and the intensity $i_1$
are shown in Fig.~\ref{fig5}.

For simplicity of the numerical analysis, we have reduced the 3D
problem to a 2D one by integrating Eq.~\ref{advection-diffusion}
along $y$. Then we obtain for
$\overline{c}=\int_{-\infty}^{\infty} c \, dy$ an equation of the
same form, with now $\nabla^2=\partial_{x^2}^2+\partial_{z^2}^2$.
The initial condition also is similar. The non-penetration
condition $\partial_z c=0$ applies automatically at the wall as a
result of the repulsive electrostatic force. The equation
advection-diffusion with these initial and boundary conditions was
solved with a commercial finite elements software (COMSOL).
Experimental values of the shear rate were used. The calculation
was repeated for several values of $b$ in the interval from zero
to 100 nm and for several values of the distance of the foci to
the wall. In each case, the calculated $\tau_M$ provided the
apparent transfer velocity between foci, $\Delta s/\tau_M$.

Our calculations allow one to reproduce both the velocity profile close to the wall and the plateau region
(see the solid dots in the inset of Fig.~\ref{fig3}). By superimposing these two parts of the
velocity profile for a given concentration of salt, we
unambiguously determine a position of the wall in the experiment.
 An apparent slip length was then
obtained by fitting a straight line through the points of the
velocity in term of the distance to the wall, similarly to the
inset in Fig.~\ref{fig3}. The apparent slip lengths is about 445
nm for all electrolyte solutions, but is much larger in pure
water, being of the order of 2 $\mu$m. These results slightly
differ from previously reported~\cite{lumma.d:2003} due to a
different, not supported by our current model, way of
determination of the wall position. As become evident from
Fig.~\ref{fig3} the fit of experimental data always gives the
values of the apparent slip close to predicted by the theory.
Thus, we conclude that the large observed value of the apparent
slip at the hydrophilic wall are fully attributed to a Taylor
dispersion affected by electrostatic interaction of nanotracers
with the wall.

The velocity profiles calculated for a slip wall are consistent
with data obtained near hydrophobic surfaces, and the apparent
slip at the hydrophobic wall is found to be $60-70$ nm larger than
in case of a hydrophilic wall  for all salt concentrations. It
follows from our modelling that the contributions of the Taylor
dispersion for the no-slip and slip are of the same order, so that
the difference of the apparent slip lengths is close to the actual
ones. Alternatively, the true hydrophobic slip length was
determined by comparison the apparent velocities at plateau
regions (see Fig.~\ref{plateau}). This procedure is more accurate
since it does not suffer from possible errors introduced by
fitting and even does not depend of the choice of the wall
location. Fig.~\ref{plateau} shows that the results for
hydrophobic surfaces are always below the curve computed for
$b=100$ nm. Note that the presence of electrolyte has little
effect on the value of the hydrophobic slip length. Neither found
we a dependence of measured values on the shear rate. In the
numerical and experimental examples we used here as an
illustration of our approach the shear rate close to the wall
(1750 s$^{-1}$) was larger than
in~\cite{tretheway.dc:2002,josef.p:2005} and comparable
to~\cite{huang.p:2006}. There are indications that shear rate
strongly influence the value of apparent slip in all range of
shear rates we used, but the true hydrophobic slip however remains
the same as in the discussed data.

In summary, we have performed an experimental study of a flow of
water-electrolyte solutions in microchanels by using a new
velocimetry technique. Our experiment is in favor of no-slip
boundary conditions for a hydrophilic
surface~\cite{vinogradova:03,charlaix.e:2005,joly.l:2006,vinogradova.oi:2006}.
It is very unlikely that there exists some minimal slip over
hydrophilic surfaces as suggested
before~\cite{huang.p:2006,josef.p:2005}. We have also demonstrated
that there is no possibility that flow exhibits a hydrophobic slip
length larger than 80-100 nm. Therefore, the slip effect is not as
extreme as many authors have
reported~\cite{tretheway.dc:2002,lauga.e:2005}. However, it is
quite large if we consider a simple molecular model for slip at
our contact angle~\cite{barrat:99}, which might be a good
indication to a two-layer
model~\cite{vinogradova:95,andrienko.d:2003} ($b/\delta\approx
\mu/\mu_g \approx 50$, where $\delta$ is the thickness of the
adjacent to the hydrophobic surface gas layer with viscosity
$\mu_g$, which suggests $\delta\approx1-2$ nm). Essentially, our
DF FCS approach allowed us to use very small particles, to reach a
very large shear rate, and to reduce dramatically an error in
measurements due to orders of magnitude better statistics than
known methods~\cite{note1}. We thus believe our Letter concludes
the discussion about boundary conditions at hydrophilic and
hydrophobic surfaces.

This work was supported by a DFG through its priority program
``Micro and Nanofluidics'' (Vi 243/1-3).  V.Lobaskin, P.Tabeling,
and R.Tsekov are thanked for discussions.

\bibliography{koynov}
\end{document}